\documentstyle[12pt]{article}

\begin{document}
\title{ Generalized intelligent states of  the $su(N)$ algebra}
\author{{\bf M. Daoud}{\footnote{Permanent adress: LPMC, Faculty of Sciences, University Ibn Zohr, Agadir,
 Morocco}}   \\
\\
Max Planck Institute for the Physics of Complex Systems\\ Dresden, Germany\
\\}

\maketitle

\begin{abstract}
Schr\" odinger-Robertson uncertainty relation is minimized for the quadrature components
of Weyl generators of the algebra $su(N)$. This is done by determining explicit Fock-Bargamann
representation of the  $su(N)$ coherent states and the differential realizations of the elements of $su(N)$.
 New classes of coherent and squeezed states are explicitly
derived.

\end{abstract}

\vfill
\newpage 

\section{Introduction}
Coherent states [1] and squeezed states [2-4] are usually associated to the minimization of the Heisenberg
uncertainty relation. However, it was proven that a relation more accruate should be used to minimize the
fluctuations of two observables when their commutator is not a multiple of unity [5-8]. This relation, known as
Schr\" odinger-Robertson uncertainty inequality [9-10], gives the so-called generalized intelligent states (see the
 pioneering works [5-7]). Following this new way to generalize the usual coherent states, there has been
  much interest for generalized intelligent states for the quadrature components corresponding to
   the generators of $su(2)$ and $su(1,1)$  algebras [8, 11-13]. They were also defined for exactly solvable
   quantum systems as eigenstate of complex combination of lowering and raising operators [14-17].\\
In this work, we extend this study to higher symmetries by developing an analytiacl approach
 that provides the generalized intelligent states for $su(N)$ algebra.
The approach is based on the Fock-Bargmann representation of $su(N)$ coherent states. The analytic representation
 , presented here, enables us to convert the algebraic
eigenvalue equations, arising from the Schr\" odinger-Robertson uncertainty relation,  into
quasi-linear differential equations. Solving these equations, one to obtain the explicit forms of the
needed intelligent
states. \\
The letter is organised as follows. Keeping in mind the utility of  the analytic reprenstation of coherent
states in determining the intelligent states, we first give the explicit expression of $su(N)$ coherent states.
This construction is based on the bosonic realization of the algebra $su(N)$. We introduce
the Fock-Bargmann space of entire analytic functions which gives a framework to simplify and to "minimize" the
problem of the derivation of intelligent states.
We give the differential actions of the $su(N)$ elements on this space.
In section 3, the analytic representation, thus constructed, are used to derive the states
minimizing the Schr\" odinger-Robertson relation for the quadrature components of Weyl elements of
$su(N)$ algebra. The
advantage of the analytic approach is clearly established. New classes of coherent  and squeezed
, as it will be explained,  emerge.  In the last section,
remarks and a number of interesting open problems are enumerated \\
\section{Analytic representations of $su(N)$-coherent states}

The algebra $su(N)$ is defined by the generators $e_i$, $f_i$, $h_i$ ($i = 1, 2,\dots, N-1$) and the relations
\begin{equation}
[ e_i , f_j ] = \delta_{ij} h_j
\end{equation}
\begin{equation}
[ h_i , e_j ] = a_{ij} e_j {\hskip 0.5cm} [ h_i , f_j ] = - a_{ij} f_j
\end{equation}
\begin{equation}
[ e_i , e_j ] = 0 \qquad \textrm{for} \qquad \vert i - j \vert > 1
\end{equation}
\begin{equation}
e_i^2e_{i \pm 1} - 2 e_ie_{i \pm 1}e_i + e_{i \pm 1}e_i^2 = 0
\end{equation}
\begin{equation}
f_i^2f_{i \pm 1} - 2 f_if_{i \pm 1}f_i + f_{i \pm 1}f_i^2 = 0
\end{equation}
where $(a_{ij})_{i,j=1,2,\dots,N-1}$ is the Cartan matrix of $su(N)$, i.e. $a_{ii} = 2$, $a_{i,i\pm 1} = -1$
and $a_{ij} = 0$ for
$\vert i-j \vert > 1$. Many aspects of Lie algebras are best considered after choosing a special type
 of the representation basis. Since one would write down the $su(N)$ coherent states, the most convenient choice
 is the bosonic realization. Indeed, an adapted basis is given in term of
  $N$ bosonic pairs of creation and annihilation operators; They satisfy the commutation relations
\begin{equation}
[ a_k^{-} , a_l^{+}] = \delta_{kl}
\end{equation}
where $k,l = 1, 2, .., N$. The occupation numbers are $ a_k^+a_k^-$. The Fock space is generated by the eigenstates
$\vert n_1, n_2,.., n_N\rangle$ of number operators, namely,
\begin{equation}
\vert n_1, n_2,\dots, n_N\rangle = \frac{(a_1^+)^{n_1}}{\sqrt{n_1!}}\frac{(a_2^+)^{n_2}}{\sqrt{n_2!}}\dots
\frac{(a_N^+)^{n_N}}{\sqrt{n_N!}}
\vert 0, 0,\cdots , 0\rangle
\end{equation}
In this bosonic representation, we define the generators of $su(N)$ as
\begin{equation}
e_i = a_i^+ a_{i+1}^- {\hskip 0.5cm} f_i = a_i^- a_{i+1}^+ {\hskip 0.5cm} h_i = a_i^+ a_{i}^- - a_{i+1}^+ a_{i+1}^-
\end{equation}
The generators $e_i$, $f_i$ are called step, ladder or Weyl operators. The Cartan subalgebra is generated by
the elements $h_i$.
They act on the representation space of dimension $\frac{(N+j_1)!}{j_1!N!}$ that is obtained from the Fock space
 , generated by eigenvectors (7),
by restricting the total number of quantas to $j_1 = n_1 + n_2 +\dots+ n_N$. In the present representation the state
of highest weight is $\vert j_1, 0,\dots, 0\rangle$. The generators of $su(N)$ having a nontrivial action
( non-vanishing and non-diagonal) on the fudicial vector
$\vert j_1, 0,\dots, 0\rangle$ are
\begin{equation}
F_2 \equiv f_1 {\hskip 0.5cm} F_i = [ f_{i-1} , F_{i-1}]
\end{equation}
for $i = 3,4,\dots, N$ and $E_i = F_i^{\dagger}$. At this stage, one can define
the coherent state as
\begin{equation}
\vert z_1 , z_2,\dots,z_{N-1} \rangle = D(z_1 , z_2,\dots,z_{N-1}) \vert j_1, 0,\dots, 0\rangle
\end{equation}
where the dispalcement operator is
\begin{equation}
 D(z_1 , z_2,\dots,z_{N-1}) = \exp \Big( \sum_{i=1}^{N-1} (z_i F_{i+1} - \bar z_i E_{i+1})\Big).
\end{equation}
Expanding the  operator $D(z_1 , z_2,\dots,z_{N-1})$ and using the actions of creation and annihilation
operators on the restricted Fock space \\
${\cal F} = \{\vert n_1, n_2,\dots,n_{N-1} \rangle ; n_1+n_2+\dots+n_{N-1} = j_1\}$, one get
\begin{eqnarray}
\lefteqn{\vert z_1 , z_2,\dots, z_{N-1} \rangle = \sum_{j_2=0}^{j_1} \sum_{j_3=0}^{j_2}\dots\sum_{j_{N-1}=0}^{j_N}
z_1^{j_2} z_2^{j_3} \dots z_{N-1}^{j_{N}}}\nonumber \\
& & {} \times I_{j_2}^{j_1}(|z_1|)I_{j_3}^{j_2}(|z_2|)\dots I_{j_{N}}^{j_{N-1}}(|z_{N-1}|)
\vert j_1-j_2, j_2-j_3,\dots , j_N\rangle.
\end{eqnarray}
where
\begin{equation}
I_{j_{s+1}}^{j_{s}}(|z_s|) = \sum_{k=0}^{\infty}\frac{(-)^k(|z_s|^2)^k}{(j_{s+1}+2k)!}
P(j_{s+1}+1,k),
\end{equation}
for $s = 1, 2,\dots,N-1$. The quantities $P$ occuring in (13) are give by
\begin{equation}
P(j_{s+1}+1,k) = P(j_{s+1}+1,0)\sum_{l_{1}=1}^{j_{s+1}+1}
p_s(l_1)\sum_{l_{2}=1}^{l_1+1}
p_s(l_2)\dots\sum_{l_{k}=1}^{l_{k-1}+1}
p_s(l_k)
\end{equation}
with $P(j_{s+1}+1,0) = \frac{j_s!j_{s+1}!}{(j_s - j_{s-1})!}$ and $p_s(l) = (j_s - l + 1)l$.
 They satisfy the following recursion formula
\begin{equation}
P(j_{s+1}+1,k) = \sqrt{p_s(j_{s+1})} P(j_{s+1},k) + \sqrt{p_s(j_{s+1}+1)} P(j_{s+1}+2,k-1).
\end{equation}
Setting
\begin{equation}
J^{j_s}_{j_{s+1}}(|z_s|) = |z_s|^{j_s}P(j_{s+1}+1,0) I^{j_s}_{j_{s+1}}(|z_s|),
\end{equation}
we get the first order differential equation
\begin{equation}
\frac{dJ^{j_s}_{j_{s+1}}(|z_s|)}{d|z_s|} = J^{j_s}_{j_{s+1}-1}(|z_s|) - (p_s(j_{s+1}+1))^2 J^{j_s}_{j_{s+1}+1}(|z_s|).
\end{equation}
The solution of this equation takes the simple form
\begin{equation}
J^{j_s}_{j_{s+1}}(|z_s|) = \frac{1}{j_{s+1}!} (\ cos (|z_s|))^{j_{s+1}-1}(\ tg (|z_s|))^{j_{s+1}},
\end{equation}
and the $su(N)$ coherent states rewrite as
\begin{eqnarray}
\lefteqn{\vert \zeta_1 , \zeta_2, \cdots , \zeta _{N-1}\rangle = {\cal N}
\sum_{j_2=0}^{j_1} \sqrt{\frac{j_1!}{j_2! (j_1 - j_2)!}}
\zeta_1^{j_2}  \sum_{j_3=0}^{j_2} \sqrt{\frac{j_2!}{j_3! (j_2 - j_3)!}} \zeta_2^{j_3}\cdots} \nonumber\\
& & {}\times
 \sum_{j_N=0}^{j_{N-1}} \sqrt{\frac{j_{N-1}!}{j_N! (j_{N-1} - j_N)!}} \zeta_{N-1}^{j_{N}}
\vert j_1 - j_2, j_2 - j_3,\cdots , j_N\rangle
\end{eqnarray}
where the normalisation constant is given by
\begin{equation}
{\cal N} = (1 + |\zeta_1|^2 + |\zeta_1|^2|\zeta_2|^2 + \cdots +|\zeta_1|^2|\zeta_2|^2  \cdots
|\zeta _{N-1}|^2)^{-\frac{j_1}{2}}
\end{equation}
and the new variables $\zeta_s$ are defined by
$\zeta_s = \frac{z_s}{|z_s|} \ tg (|z_s|) \cos (|z_{s+1}|)$ for $s= 1, 2, \cdots , N-2$ and
 $\zeta_{N-1} = \frac{z_{N-1}}{|z_{N-1}|} \ tg (|z_{N-1}|)$.
 It is important to note that the coherent states (19) can be obtained also from Eq.(3.31) of the work [18],
  but it is necessary to follows the procedure, based on the equations (12-18), presented in this section.
  In other words, the authors of [18] have avoided the explicit computation of the action of the displacement
  operator on the highest weight state $\vert j_1, 0, \cdots, 0\rangle$. It is evident that for $N=2$, one
  recover the well known $su(2)$ coherent states. The states (19) have
the property of strong continuity in the label space and overcompletion in the sense
that there exists a positive measure such that they solve the resolution to identity.
 The appropriate form of this resolution is
\begin{eqnarray}
\lefteqn{\int d\mu
\vert \zeta_1, \zeta_2,\cdots , \zeta_{N-1}\rangle \langle \zeta_1 , \zeta_2, \cdots , \zeta_{N-1}\vert =
\sum_{j_2=0}^{j_1} \sum_{j_3=0}^{j_2} \cdots \sum_{j_N=0}^{j_{N-1}}}\nonumber\\
& & \vert j_1 - j_2, j_2 - j_3, \cdots , j_N\rangle
  \langle j_1 - j_2, j_2 - j_3, \cdots, j_N \vert.
\end{eqnarray}
Assuming the isotropy of the measure $d\mu $,  we set
\begin{equation}
d\mu  = \pi^{N-1} {\cal N}^{-1}
 \prod_{s=1}^{N-1}h( \vert \zeta_s \vert ^2)
 \vert \zeta_s\vert d \vert \zeta_s\vert
 \vert d\theta_s
\end{equation}
with $\zeta_s = \vert \zeta_s \vert e^{i\theta_s}$.
Substituting (22) in Eq.(21), we obtain the following sum
\begin{equation}
\int_{0}^{\infty} x_s^{j_{s+1}} h(x_s) dx_s = \frac{j_{s+1}!(j_{s}-j_{s+1})!}{j_s!}.
\end{equation}
which should be satisfied
by the function $h(x_s = \vert \zeta_s \vert ^2 ))$.
One get
\begin{equation}
h(x_s) = \frac{j_{s}+1}{(1+x_s^2)^{j_s+2}}.
\end{equation}
This result can be obtained by using the definition of Meijer's $G$-function and
the Mellin inversion theorem [19]. The resolution to identity is necessary to build up the Fock-Bargamann
space based on the set of $su(N)$ coherent states.\\

It is well established that  the use of the Fock-Bargmann representation is a powerful method for
obtaining closed analytic expressions for various properties of coherent states. Calculation for some
quantum exceptation values and solutions for some eigenvalue equations are simlpified by exploiting the theory
of analytical entire functions. Here, we give the Fock-Bargamnn representation for a quantum
system whose its dynamical symmetry is described by the Lie algebra $su(N)$.
 We define the Fock-bargamnn space as a space of functions which are holomorphic. The scalar product
is written with an integral of the form
\begin{equation}
\langle f \vert g \rangle = \int {\bar f(\zeta _1, \zeta _2, \cdots ,\zeta _{N-1} )}
g(\zeta _1, \zeta _2, \cdots , \zeta _{N-1}) d\mu
\end{equation}
where the measure is defined above (see Eqs.(22) and (24)). Due to overcompletion of the coherent sates,
it is induced by the scalar
product in ${\cal F}$. Let
\begin{equation}
\vert \psi \rangle = \sum _{n_1+n_2+ \cdots +n_N = j_1} a_{n_1,n_2, \cdots ,n_N}\vert n_1,n_2, \cdots ,n_N \rangle
\end{equation}
an arbitrary quantum state of ${\cal F}$, it can be represented
as a function of the complex variables $\zeta_1, \zeta_2, \cdots ,\zeta _{N-1}$ as
\begin{equation}
\psi (\zeta_1, \zeta_2, \cdots ,\zeta _{N-1}) =
{\cal N}^{-1}
\langle \bar \zeta_1, \bar \zeta_2 , \cdots ,\bar \zeta _{N-1}\vert \psi \rangle
\end{equation}
 In particular, the analytic functions associated to elements of the basis of ${\cal F}$ are defined as
\begin{equation}
\psi _{j_1, j_2,\cdots,j_N}(\zeta_1, \zeta_2,\cdots ,\zeta _{N-1}) = {\cal N}^{-1}
\langle \bar \zeta_1, \bar \zeta_2, \cdots ,\bar \zeta _{N-1}
\vert j_1 - j_2, j_2 - j_3,\cdots, j_N \rangle
\end{equation}

We now investigate the form of the action of the operators $e_i$, $f_i$ and $h_i$ on Fock-Bargmann space
generated by the functions (28).\\
Indeed, any operator $O$ of the algebra $su(N)$ is represented in the space of entire analytical functions by some differential operator
${\cal O}$, defined by
\begin{equation}
\langle \bar \zeta_1, \bar \zeta_2, \cdots ,\bar \zeta _{N-1} \vert O \vert \psi \rangle
= {\cal O} \psi (\zeta_1, \zeta_2, \cdots ,\zeta _{N-1})
\end{equation}
for any state $\vert \psi \rangle $ of ${\cal F}$.\\
According this definition, we obtain
\begin{equation}
E_{i+1} = \partial _i {\hskip 0.5cm} F_{i+1} = j_1 \zeta_i - \zeta_i^2 \partial _i
- \zeta_i \sum_{i\neq k}\zeta_k \partial _k
\end{equation}
for $i=1,2, \cdots, N-1$ and where $\partial _i$ stands for the derivative in respect to the variable $\zeta _i$.
To obtain the above differential realization:\\
\indent (i) we remark that the coherent states (19) can be also written as
\begin{equation}
\vert \zeta _1, \zeta _2, \cdots , \zeta _{N-1}\rangle =
{\cal N}
D(\zeta _1, \zeta _2, \cdots ,\zeta _{N-1}) \vert j_1 , 0,\cdots , 0 \rangle
\end{equation}
where $D(\zeta _1, \zeta _2, \cdots ,\zeta _{N-1}) = \exp (\sum_{i=1}^{N-1}\zeta _iF_{i+1} )$,\\
\indent (ii) we observe that
\begin{equation}
\partial _i D(\zeta _1, \zeta _2, \cdots ,\zeta _{N-1}) = F_{i+1} D(\zeta _1, \zeta _2, \cdots ,\zeta _{N-1})
\end{equation}
\indent (iii) we use the Hausdorff formula
\begin{equation}
e^{-B} A e^B = \sum _{n \geq 0} \frac{1}{n!} \big( - adB\big)^n A
\end{equation}
where $\big( adB\big) A = [ B, A]$, \\
\indent (iv) we use also the actions of the elements of $su(N)$ on the basis of Fock space ${\cal F}$, in particular the
fudicial vector $\vert j_1, 0, \cdots , 0 \rangle$, and the structure relations (1-5) of the algebra $su(N)$.\\
It follows that the elements $e_i$, $f_i$ and $h_i$ of the algebra $su(N)$ are realized as
\begin{equation}
f_1 = j_1 \zeta_1 - \zeta_1^2 \partial _1
- \zeta_1 \sum_{k = 2}^{N-1}\zeta_k \partial _k,  {\hskip 0.3cm}  e_1 = \partial _1,  {\hskip 0.3cm} h_1 =
j_1 - 2 \zeta_1\partial _1 - \sum_{k=2}^{N-1} \zeta _k \partial _k,
\end{equation}
\begin{equation}
e_i = \zeta_{i+1}\partial _i, {\hskip 0.3cm}  f_i = \zeta_i\partial _{i+1} ,{\hskip 0.2cm}
h_i = \zeta_i\partial _i - \zeta_{i+1}\partial _{i+1},
\end{equation}
for $i= 2, 3, \cdots, N-1$.
Hence, as it is clear from the previous considerations, the $su(N)$ generators  act as
 first-order holomorphic differential operators on the Fock-Bargmann space generated by the
 elements (28). One can verify that the
commutation relations (1-5) are preserved. This result combined with eigenvalue equations
ensuring the minimization
of Schr\" odinger-Robertson inequality provides the intelligent states
as that will be explained in the next section.

\section{\bf Robertson states for $su(N)$ Weyl generators}
In this section, we will study
 the fluctuations of quadrature components of Weyl generators which represent creation and annihilation
 of states for a quantum mechanical system of $su(N)$ symmetry. In this order, to construct the intelligent
 states of any pair of ladder operators $e_i$, $f_i$ ($i=1, 2, \cdots, N-1$), it is natural to introduce
 the quantum observables  $ \sqrt{2}p_i = e_i + f_i$ and
${\bf i}\sqrt{2}q_i = e_i - f_i$  where ${\bf i}^2=-1$. These observables obey
\begin{equation}
[p_i , q_i ] = {\bf i} h_i
\end{equation}
 We known that $p_i$ and $q_i$ satisfy, in a given state, the Robertson-Shr\" odinger
uncertainty
relation
\begin{equation}
(\Delta p_i)^2 (\Delta q_i)^2 \geq \frac{1}{4} ( \langle h_i\rangle^2 + \langle c_i\rangle^2)
\end{equation}
where $\Delta p_i$ and $\Delta q_i$ are the dispersions and the hermitian operator$c_i = \{ p_i - \langle p_i\rangle, q_i - \langle q_i\rangle\}$
gives the covariance (correlation) of the observables $p_i$ and $q_i$. The symbol $\{ , \}$ stands for the standard definition of the
anticommutator. A state $\vert \Phi \rangle$ providing the equality in (37) is
the so-called generalized intelligent state. It was proven that such state satisfy the following
eigenvalue equation
\begin{equation}
\big((1 + \alpha) e_i + (1 - \alpha) f_i \big) \vert \Phi \rangle =  \lambda \vert \Phi \rangle
\end{equation}
where $\alpha \neq 0$ and $\lambda = (1 + \alpha) \langle e_i\rangle + (1 - \alpha) \langle f_i\rangle$
are complex parameters. Furthermore, the variance and covariance, in the intelligent state
$\vert \Phi \rangle$, are given by
\begin{equation}
(\Delta p_i)^2 = \vert \alpha \vert \Delta _i  {\hskip 0.5cm}  (\Delta q_i)^2 = \frac{1}{\vert \alpha \vert} \Delta _i
\end{equation}
where $ \Delta _i = \frac{1}{2} \sqrt{ \langle h_i\rangle^2 + \langle c_i\rangle^2}$.
Remark that they can be also expressed as
\begin{equation}
(\Delta p_i)^2 = \frac{\vert \alpha \vert^2}{u}\langle h_i\rangle {\hskip 0.5cm}
(\Delta q_i)^2 = \frac{1}{u}\langle h_i\rangle {\hskip 0.5cm}
\langle c_i\rangle = \frac{v}{u} \langle h_i\rangle
\end{equation}
where the real parameters $u$ and $v$ are such that $u^2 + v^2 = 4 \vert \alpha \vert^2$ ( As example, one can take $ u = 2Re\alpha$ and $v = 2Im\alpha$
). It is clear that the dispersions
and the correlation can be obtained from the mean value of the observable $h_i$.
The state $\vert \Phi \rangle $ satisfying
 (38) with $\vert \alpha \vert = 1$ are coherent because they satisfy $(\Delta p_i)^2
  = (\Delta q_i)^2 = \Delta _i$. The fluctuations are equals and minimized
  in the sense of Schr\" odinger-Robertson uncertainty relation. The state satisfying
  (38) with $\vert \alpha \vert \neq 1$ are squeezed because if $\vert \alpha \vert  < 1$,
  we have $(\Delta p_i)^2 < \Delta _i < (\Delta q_i)^2$ and for $\vert \alpha \vert > 1$,
  we have $(\Delta q_i)^2 < \Delta _i < (\Delta p_i)^2$.
  \\
\indent To solve the eigenvalues equation (38), we will use the anlytic representations of coherent
 states as well as
the differential realizations of the the generators $e_i$ and \\$f_i$ .
 So, let us start by deriving the eigenfunctions of Eq.(38) for the first pair  $e_1$, $f_1$.
 By introducing the analytic function
 \begin{equation}
 \Phi _1 \equiv \Phi _1( \zeta _1,\zeta _2,\cdots, \zeta _{N-1}, \alpha , \lambda , j_1) =
  {\cal N}^{-1}
\langle \bar \zeta_1, \bar \zeta_2, \cdots, \bar \zeta_{N-1} \vert \Phi _1 \rangle,
 \end{equation}
 it can be easly checked that the eigenvalue equation (38) can be converted in the following first order
  differential equation
\begin{equation}
(j_1 \eta_1 - \lambda ') \Phi _1 + ( 1 - \eta _1^2 )\frac{\partial \Phi _1}{\partial \eta_1}
- \eta _1 \sum_{i=2}^{N-1}\eta_i \frac{\partial \Phi _1}{\partial \eta_i}  = 0,
\end{equation}
where $ \eta _1 = \sqrt{\frac{1-\alpha}{1+\alpha}} \zeta_1$, $\eta _{i \neq 1} = \zeta_i$ and
 $\lambda'= \frac{\lambda}{\sqrt{1-\alpha ^2}}$
for $ \alpha \neq \pm 1 $. The function
$ \Phi _1( \zeta _1,\zeta _2, \cdots, \zeta_{N-1} \alpha , \lambda , j_1)$  and can be expanded as
\begin{equation}
\Phi _1 = \sum _{j_2=0}^{j_1}\sum_{j_3=0}^{j_2} \cdots \sum_{j_N=0}^{j_{N-1}}
 a_{j_1,j_2,\cdots, j_N} \eta _1^{j_2} \eta _2^{j_3}\cdots  \eta _{N-1}^{j_N}
\end{equation}
Substitution of (43) in (42) yields the recursion formula
\begin{equation}
(j_1 + 1 - \sum_{i=2}^{N}j_i) a_{j_1,j_2 - 1, \cdots,j_N} - \lambda ' a_{j_1,j_2,\cdots, j_N} +
(j_2 + 1) a_{j_1,j_2 + 1, \cdots,j_N} = 0
\end{equation}
which  can be solved
by the Laplace method. Indeed, we set
\begin{equation}
a_{j_1,j_2,\cdots, j_N} = \int_{-1}^{+1} x^{j_2} f(x) dx
\end{equation}
that we introduce in (44) to obtain, after partial integration, the simple first order differential equation
satisfied by the function $f(x)$
\begin{equation}
(x - x^3) \frac{df}{dx} +  (2j + 1 - \lambda 'x - x^2) = 0.
\end{equation}
where $2j = j_1 - \sum_{i=3}^{N}j_i$. The last equation is easly solvable. Replacing in (45), one get
\begin{equation}
a_{j_1,j_2,\cdots, j_N} =
 \int _{-1}^{+1} x^{j_2-2j-1} (1-x)^{-\frac{\lambda'}{2}+ j}(1+x)^{\frac{\lambda'}{2}+j}dx,
\end{equation}
or
\begin{eqnarray}
\lefteqn{a_{j_1,j_2,\cdots,j_N} = (-)^{j_2} \frac{\Gamma (\frac{\lambda'}{2}+j+ 1)
\Gamma (-\frac{\lambda'}{2}+j+ 1)}
{\Gamma (2j + 2)}} \nonumber\\
& &{} \times {_2}F_1(2j-j_2+1, \frac{\lambda'}{2}+j+ 1, 2j+2, 2)
\end{eqnarray}
using the integral representation for the hypergeometric function $_2F_1$ [19] with the condition
$-(j+1) < Re(\lambda'/2) < (j+1)$.
 Comparing the expansion (43)
 with the general
formula (41), we have the decomposition of the intelligent states over the basis of Fock space ${\cal F}$
\begin{eqnarray}
\lefteqn{\vert \Phi _1 \rangle = \sum _{j_2=0}^{j_1}\sum_{j_3=0}^{j_2}\dots\sum_{j_N=0}^{j_{N-1}}
a_{j_1,j_2,\cdots,j_N}
\Big(\frac{1-\alpha}{1+\alpha}\Big)^{\frac{j_2}{2}}
\sqrt{\frac{j_N!}{j_1!}}}\nonumber\\
& & {} \sqrt{(j_1-j_2)!(j_2-j_3)!\cdots (j_{N-1}-j_N)!} \vert j_1-j_2, j_2-j_3,\cdots, j_N\rangle
\end{eqnarray}
where the coefficients $a_{j_1,j_2,\dots,j_N}$ are given by Eq.(48).\\
\indent Now we consider the construction of intelligent states for the  pairs  $e_i$,$f_j$
with $i = 2, 3, \cdots, N-1$ .
The eigenvalues equation (38)
 gives, in this case, the following quasi linear differential equation
\begin{equation}
\xi_i \frac{\partial \Phi_i}{\partial \xi_{i+1}} +
\xi_{i+1} \frac{\partial \Phi_i}{\partial \xi_i} - \lambda'\Phi_i = 0
\end{equation}
where $\xi_i = \sqrt{\frac{1+\alpha}{1-\alpha}}\zeta_i$, $\xi_{i+1} = \zeta_{i+1}$
and $\lambda' = \frac{\lambda}{\sqrt{1-\alpha^2}}$.\\
Here also, we expand the eigenfunction $\Phi_i \equiv
\Phi_i( \xi _1,\xi _2,\cdots, \xi _{N-1} ,\alpha , \lambda , j_1)$ as
\begin{equation}
\Phi_i = \sum_{j_2=0}^{j_1}\sum_{j_3=0}^{j_2}\cdots \sum_{j_N=0}^{j_{N-1}}
 b_{j_1,j_2,\cdots, j_N}\xi_1^{j_2}\xi_2^{j_3}\cdots \xi_{N-1}^{j_N}
\end{equation}
that we insert in the equation (50) to obtain the recursion relation linking the coefficients $b$'s
\begin{equation}
(j_{i+2} + 1)b_{j_1,\dots,j_{i+1}-1,j_{i+2}+1,\cdots,j_N} -
\lambda'b_{j_1,\cdots,j_{i+1},j_{i+2},\cdots,j_N} +
(j_{i+1} + 1) b_{j_1,\cdots,j_{i+1}+1,j_{i+2}-1,\cdots,j_N} = 0
\end{equation}
Setting $ b_{j_1,\dots,j_{i+1},j_{i+2},\cdots,j_N} \equiv b_{j_{i+1},j_{i+2}}
\equiv b_{j_{i+1}- l, l}$ where $ 2l = j_{i+1}+j_{i+2}$, the previous relation can be transformed to
\begin{equation}
(j_{i+2} + 1) b_{j_{i+1}- l-1, l} - \lambda'b_{j_{i+1}- l, l} + (j_{i+1}  + 1)b_{j_{i+1}- l+1, l} = 0,
\end{equation}
sovable in a similar manner that one given the solution of recursion formula (44), and one has
\begin{eqnarray}
\lefteqn{b_{j_1,j_2,\cdots,j_N} = (-)^{j_{i+1}} \frac{\Gamma (\frac{\lambda'}{2}+l+ 1)
\Gamma (-\frac{\lambda'}{2}+l+ 1)}{\Gamma (2l + 2)}} \nonumber\\
& & {} \times {_2}F_1(2l-j_{i+1}+1, \frac{\lambda'}{2}+l+ 1, 2l+2, 2)
\end{eqnarray}
where $-(l+1) < Re(\lambda'/2) < (l+1)$. Finally, one obtain
\begin{eqnarray}
\lefteqn{\vert \Phi _i \rangle = \sum _{j_2=0}^{j_1}\sum_{j_3=0}^{j_2}\dots\sum_{j_N=0}^{j_{N-1}}
b_{j_1,j_2,\cdots,j_N}
\Big(\frac{1+\alpha}{1-\alpha}\Big)^{\frac{j_2}{2}}
\sqrt{\frac{j_N!}{j_1!}}}\nonumber\\
& & {} \sqrt{(j_1-j_2)!(j_2-j_3)!\cdots (j_{N-1}-j_N)!} \vert j_1-j_2, j_2-j_3,\cdots, j_N\rangle
\end{eqnarray}
From equations (38) and (39), it is clear that the intelligent states $\vert \Phi_i \rangle$,
 ($i=1,2,\cdots,N-1$,) given by (49) and (55) are coherent for $\alpha = e^{{\bf i}\theta}$
 ($\theta$ real) and
 squeezed for $\vert \alpha \vert \neq 1$ in the sense of Schr\" odinger-Robertson uncertainty relation.
To close this section, let us also note  that the Fock-Bargmann representation of
the coherent states plays a helpful role in the problem of finding intelligent
states of the quadrature of Weyl generators.
 The procedure descibed here can be relevant in the determination of intelligent states for
quadrature components of type $e_i$, $e_j$ and $f_i$, $f_j$ ($i \neq j$).

\section{\bf Summary and outlook}
We constructed explicitly the coherent states associated with the Lie algebra of type $su(N)$. We have proceeded,
in a second stage, in the study of Fock-Bargmann representation based on the obtained states. We gave the
differential actions of $su(N)$ generators on this space. We have shown that they act as first order differential
operators. As byproducts, simple quasi-linear equations, satisfied by the minimum Schr\" odinger-Robertson
uncertainty states, are solved. Thus, new classes of coherent $(\vert \alpha \vert = 1)$ and squeezed
$(\vert \alpha \vert \neq 1)$ states are obtained for the quadrature components of $su(N)$ Weyl generators. It is clear
that the analytic approach used through this work can be applied for finding Robertson intelligent
 states associated to the other quadratures of the $su(N)$ generators. They can be also obtained
 by considering the eigenvalue problem for an operator which is a complex linear combination
 of all elements of $su(N)$
 \begin{equation}
 \sum_{i=1}^{N-1} ( \alpha_i^+ e_i + \alpha_i^- f_i + \alpha_i^0 h_i) \vert \Psi \rangle = \lambda
 \vert \Psi \rangle.
 \end{equation}

\noindent The solutions of such general problem give the so-called in the literature algebra
 eigenstates or algebraic coherent states ([11-13] and references therein). Taking specific constraints
 on the complex parameters
 occurring in this general eigenvalue equation, one can get various kind of coherent and
 squeezed states, in particular ones not discussed in this letter. This constitutes
 the first possible prolonogation of this work. The results of this note give the
 necessary ingredients to write down a complete classification of $A_N$ coherent and squeezed states. Indeed,
 although we have performed the analysis for the compact algebra $su(N)$, our results also describe the non-compact
  algebra $su(p,q)$ $(p + q = N)$. This can be done following the general procedure for associating a non-compact
   algebra with a compact one. Also, as continuation, it would be interesting
 to apply the approach given here to the Lie algebras  $B_N$ , $C_N$ and $D_N$.\\
{\vskip 1.0cm}

{\bf Acknowledgements}:
The author would like to acknowledge the Max Planck Institute for the Physics of Complex
Systems for kind hospitality and helpful atomosphere .\\

{\vskip 1.0cm}

\vfill\eject

\end{document}